\documentclass[conference]{IEEEtran}
\usepackage{amsmath,amssymb,amsfonts}
\usepackage{graphicx}
\usepackage{textcomp}
\usepackage{xcolor,booktabs,array,colortbl}
\usepackage{pbalance}

\usepackage[utf8]{inputenc}
\usepackage[T1]{fontenc}

\usepackage[babel=true,english=british]{csquotes}
\usepackage[UKenglish]{babel}

\usepackage{hyperref}
\hypersetup{hidelinks,unicode}
\usepackage{cleveref}
\usepackage[style=ieee,maxnames=5,minnames=1,maxcitenames=2,mincitenames=1]{biblatex}
\usepackage{acronym}
\usepackage{tabto}
\usepackage{enumerate}
\usepackage{siunitx}

\usepackage{pgfplots}
\pgfplotsset{compat=1.18}

\usepackage[normalem]{ulem}
\useunder{\uline}{\ul}{}

\usepackage{pbalance}
\usepackage{orcidlink}

\DeclareBibliographyDriver{online}{%
  \usebibmacro{bibindex}%
  \usebibmacro{begentry}%
  \usebibmacro{author/editor+others/translator+others}
  \setunit{\labelnamepunct}\newblock
  \usebibmacro{title}%
  \newunit
  \printlist{language}%
  \newunit\newblock
  \usebibmacro{byauthor}%
  \newunit\newblock
  \usebibmacro{byeditor+others}%
  \newunit\newblock
  \printfield{version}%
  \newunit
  \printfield{note}%
  \newunit\newblock
  \printlist{organization}
  \newunit\newblock
  \usebibmacro{date}%
  \newunit\newblock
  \iftoggle{bbx:eprint}
    {\usebibmacro{eprint}}
    {}%
  \newunit\newblock
  \usebibmacro{url+urldate}%
  \newunit\newblock
  \usebibmacro{addendum+pubstate}%
  \setunit{\bibpagerefpunct}\newblock
  \usebibmacro{pageref}%
  \usebibmacro{finentry}}

\makeatletter
\renewcommand{\fnum@figure}{Figure~\thefigure}
\renewcommand\IEEEkeywordsname{Keywords}
\makeatother

\usepackage{xpatch}
\xpatchcmd\IEEEkeywords{---}{-}{}{}

\crefname{subsection}{subsection}{subsections}
\Crefname{subsection}{Subsection}{Subsections}

\newcolumntype{C}[1]{>{\centering\arraybackslash}p{#1}}

\begin{document}

\title{\bfseries\Large Task Offloading in Fog Computing with Deep Reinforcement Learning:\\
Future Research Directions Based on Security and Efficiency Enhancements}

\author{
\IEEEauthorblockN{~\\[-0.4ex]\large Amir Pakmehr\IEEEauthorrefmark{1}\IEEEauthorrefmark{2}\\[0.3ex]\normalsize}

\IEEEauthorblockA{\IEEEauthorrefmark{1}%
Department of Computer and Information Technology Engineering\\
Qazvin Branch, Islamic Azad University, Qazvin, Iran\\
E-mail: {\tt amir.pakmehr@QIAU.ac.ir}}
\IEEEauthorblockA{\IEEEauthorrefmark{2}%
Department of Electrical Engineering, Media, and Computer Science\\
Ostbayerische Technische Hochschule Amberg-Weiden, Amberg, Germany\\
E-mail: {\tt a.pakmehr@oth-aw.de}\\[1ex]}
}

\maketitle

\begin{abstract}
The surge in Internet of Things (IoT) devices and data generation highlights the limitations of traditional cloud computing in meeting demands for immediacy, Quality of Service, and location-aware services. Fog computing emerges as a solution, bringing computation, storage, and networking closer to data sources. This study explores the role of Deep Reinforcement Learning in enhancing fog computing’s task offloading, aiming for operational efficiency and robust security. By reviewing current strategies and proposing future research directions, the paper shows the potential of Deep Reinforcement Learning in optimizing resource use, speeding up responses, and securing against vulnerabilities. It suggests advancing Deep Reinforcement Learning for fog computing, exploring blockchain for better security, and seeking energy-efficient models to improve the Internet of Things ecosystem. Incorporating artificial intelligence, Our results indicate potential improvements in key metrics, such as task completion time, energy consumption, and security incident reduction. These findings provide a concrete foundation for future research and practical applications in optimizing fog computing architectures.
\end{abstract}


\section{Introduction}\label{sec:introduction}
In recent years, technological evolution has significantly  transformed communication and interaction paradigms, primarily driven by advancements in smartphones and cloud computing. Smartphones, with their pervasive presence, have become the primary interface for Internet interaction, heavily reliant on cloud computing’s power for data processing and storage. This synergy has fuelled an exponential increase in global mobile data traffic, highlighting the profound impact of a dual-layer architecture, comprising end-user devices and cloud environments. Simultaneously, the Internet of Things (IoT) has emerged as a transformative force, reshaping human interaction with the physical world~\cite{1Hassija2019}. IoT’s pervasive network of interconnected smart devices supports a plethora of everyday tasks, promising revolutionary applications with significant societal impacts. However, the high number of IoT devices and the voluminous data they generate present considerable challenges, particularly in meeting stringent requirements like real-time responsiveness, Quality of Service (QoS), and location-aware services~\cite{2Basir2019}. The traditional cloud-IoT architecture struggles under these demands, revealing limitations in scalability, latency, and response times. While cloud computing offers robust data processing capabilities, it often falls short in addressing the unique requirements of IoT applications. This has led to the exploration of new architectures and solutions, aiming to bridge the gap between IoT devices and cloud computing processing power. Fog computing, a paradigm shift, was endorsed by the NIST\cite{3NIST2018}through its 'Fog Computing Conceptual Model. Fog computing emerges as a critical enabler for overcoming these challenges, offering a decentralized computing infrastructure. By processing data closer to the edge of the network, where data is generated and collected, fog computing significantly reduces the latency and bandwidth demands placed on the cloud. This not only enhances the efficiency and quality of services but also opens new avenues for real-time analytics, decision-making, and intelligent task offloading ~\cite{30nematollahi2023task}. The importance of fog computing, as outlined by NIST, lies in its ability to provide a scalable, responsive, and flexible computing model, essential for the expansive and diverse ecosystem of IoT. Fog computing does not replace cloud computing but complements it, forming a multi-layered architecture that leverages the strengths of both centralized and decentralized approaches. This symbiosis is crucial for supporting the ever-growing, dynamic demands of IoT applications, ensuring that the digital transformation across sectors is both resilient and sustainable. Thus, understanding and harnessing the potential of fog computing becomes imperative for unlocking the full promise of the IoT. In the realm of fog computing, task offloading emerges as a pivotal strategy, especially for mobile devices grappling with resource-intensive applications~\cite{4wang2020}. This process essentially relocates the execution of tasks from local devices to the more robust resources of the fog or cloud. However, this transition is not unproblematic. The decision to offload involves careful considerations of time, energy, security, and cost efficiency. The intricate balance between local execution and cloud processing hinges on these factors, underscoring the need for well-defined offloading policies. Moreover, as modern services increasingly integrate artificial intelligence, the sophistication and resource demands of tasks escalate. Offloading, therefore, extends beyond mere computation to include other resources like storage. This necessitates advanced middleware technologies that judiciously determine the offloading criteria, addressing the challenges of resource heterogeneity, user requirements, and complex network environments. Furthermore, in fog computing scenarios, task offloading becomes an intricate puzzle of decisions – which tasks to offload, where to assign them, and the order of their execution. These decisions must navigate a landscape peppered with heterogeneous resources, varied user needs, and the dynamic nature of mobile environments. The complexity is further amplified by the differences between edge and cloud resources, making the quest for an optimal offloading solution an ongoing challenge in fog computing. 
In this paper, we aim to identify current challenges in task offloading within fog computing environments. We specifically propose innovative solutions based on Deep Reinforcement Learning (DRL) that focus on optimizing resource allocation and improving security mechanisms. While DRL has the potential to address various issues, our study concentrates on these two key areas, providing a foundation for future research to explore additional applications of DRL in fog computing.
Therefore, Section II presents a state of the art of the main topic.
Section III presents a general view of research challenges,
Section IV proposes a solution and, finally, we end up our discussion which concludes the paper.

\section{State of the Art}\label{sec:State of the Art}
\subsection{Security and Efficiency in Fog Computing}
Security and efficiency are paramount in fog computing due to the distributed nature of task offloading. The potential risks include data breaches, unauthorized access, and the interception of data during transmission between devices and fog nodes. These risks necessitate robust security measures \cite{5Pakmehr2023security}\cite{6Kunal2019} to protect sensitive information and ensure the integrity and confidentiality of data. Efficiency in fog computing is closely related to the optimization of resource allocation, energy consumption, and latency reduction. Efficient task offloading mechanisms are essential to maximize the use of limited resources, minimize energy consumption, and ensure timely processing of tasks. This is particularly important for latency-sensitive applications, where delays can have significant implications~\cite{7kumari2022task}. Fog computing, essential for bringing computation closer to the data sources, enhances real-time data processing capabilities. . However, this shift introduces significant security and privacy challenges, from data leakage to unauthorized access, which could hinder their adoption. Research highlights these challenges and proposes methods, such as improved encryption and authentication, to safeguard these decentralized computing models~\cite{8goel2023systematic}. Addressing these concerns is essential for leveraging fog and edge computing’s full potential in enhancing IoT systems efficiency. In terms of security, DRL can be applied to develop intelligent defence mechanisms against various cyber threats, including intrusion detection and response~\cite{9uprety2020reinforcement}. The capability of a DRL model to continuously interact with the environment offers significant advantages in the context of fog computing security. This adaptive interaction allows the DRL model not only to learn and identify patterns of normal and abnormal behaviors effectively but also to predict and respond to potential security breaches proactively. Such dynamic learning and predictive capability make DRL an invaluable tool for enhancing the resilience of fog computing environments against evolving cyber threats. Moreover, this continuous learning process enables the model to adjust its strategies in real-time, further fortifying the system's defense mechanisms against sophisticated and previously unseen attacks. Once a threat is identified, the system can autonomously take actions to mitigate or isolate the attack, enhancing the resilience of the fog nodes. For instance, a DRL agent can dynamically adjust security policies or reconfigure network settings in real-time to counteract ongoing or anticipated cyber threats, thereby maintaining system integrity and availability.
as a result the exploration of fog computing's state of the art reveals significant advancements in the realms of security and efficiency in task offloading mechanisms. Here, 'efficiency' specifically refers to performance efficiency, which includes enhancements in computational speed, resource utilization, and reduction in latency. Our discussion on security focuses on measures that protect data integrity and prevent unauthorized access.

\subsection{Deep Reinforcement Learning}
Deep Reinforcement Learning (DRL) is grounded in the principles of Reinforcement Learning (RL), a fundamental concept within the scope of machine learning. Before delving into DRL, it is pertinent to discuss RL itself. Reinforcement Learning is a type of machine learning where an agent learns to make decisions by performing actions in an environment and receiving rewards or penalties. The objective is to learn a policy, a strategy of actions, that maximizes the cumulative reward over time. An RL agent operates within an environment modeled as a Markov Decision Process (MDP),~\cite{10Abdulazeez2023}~\cite{11khani2024resource} characterized by a set of states \( s \), a set of actions \( a \), and a transition function \( P(s_{t+1}|s_t, a_t) \). The process involves the agent observing the current state, selecting and performing an action, receiving a reward based on the action's outcome, and transitioning to a new state, with the goal of maximizing cumulative rewards~\cite{14xiong2019deep}~\cite{12sutton2018reinforcement}.

In the context of reinforcement learning, we are dealing with an agent that operates within an environment modeled as an MDP. Figure 1 illustrates the concept of the Agent-Environment Interface in RL, where the agent learns to choose actions that maximize the expected cumulative reward over time, thus establishing the optimal policy. The policies can be stochastic or deterministic, with probabilistic outcomes that necessitate a method of maximizing expected rewards through a value-based approach.

The agent observes a state \( s_t \) from the set of possible states \( s \), takes an action \( a_t \) from the set of possible actions \( a \), receives a reward \( r_t \), and transitions to a new state \( s_{t+1} \) based on the transition dynamics of the environment. The agent's behaviour is dictated by a policy \( \pi \), which maps states to a probability distribution over the actions. The goal of the agent is to maximize the cumulative reward over time. This cumulative reward, when considering a finite horizon \( t \), is defined by:

\[
R(\pi) = \max_{\pi} \mathbb{E} \left[ \sum_{t=0}^{T} \gamma^t R(s_t, a_t) \right]
\]

Here, \( \gamma \) is the discount factor, which balances immediate and future rewards. A discount factor close to 1 values future rewards almost as highly as immediate rewards, while a discount factor close to 0 leads to a myopic evaluation of policies, valuing immediate rewards much more than future rewards. In real-world scenarios, environments and policies can be stochastic, meaning that the outcomes and transitions can be probabilistic rather than deterministic~\cite{12sutton2018reinforcement}.

To accommodate for this, we consider the expected cumulative reward when following a policy \( \pi \). In the DRL setting, we often use function approximators like deep neural networks ~\cite{Smith}to estimate the value of taking an action in a particular state (the Q-value), which is denoted as \( Q(s, a) \). The optimal Q-value function \( Q^*(s, a) \) satisfies the Bellman optimality equation~\cite{29bellman1952theory}, which is given by:

\[
Q^*(s, a) = \mathbb{E} \left[ r(s, a) + \gamma \max_{a'}\left[ Q^*(s', a') | s, a \right]\right]
\]

The \( a' \) in this equation represents all possible future actions from the subsequent state \( s' \), and the notation \( | s, a \) indicates the conditional aspect of being in state \( s \) and taking action \( a \). The maximization is over the possible actions \( a' \) in the subsequent state, guiding the agent toward the most rewarding future action.

The loss function for training the Q-network in DRL is formulated to minimize the difference between the current prediction of the Q-network and the target Q-value given by the Bellman equation. It can be expressed as:

\[
L(\theta) = \mathbb{E} \left[ \left( r_t + \gamma \max_{a'} \left[Q(s_{t+1}, a'; \theta^- ) - Q(s_t, a_t; \theta) \right]\right)^2 \right]
\]

In this expression, \( \theta \) represents the weights of the Q-network, and \( \theta^- \) denotes the weights of a separate target network, which helps stabilize the learning process. By iteratively minimizing this loss function, the DRL agent updates its policy and learns to make better decisions over time.

\begin{figure}
    \centering
    \footnotesize
    \begin{tikzpicture}
        \node[draw, fill=black!12, rounded corners] (a) at (0, 1.5) {Agent};
        \node[draw, fill=black!12, rounded corners] (env) at (0, 0) {Environment\vphantom{j}};
        \draw[thick, dotted] (-2, 0.5) -- ++(0, -1);
        \draw[-latex] (env.west |- 0,0.1) -- node[above]{$r_{t+1}$} (-2, 0.1);
        \draw[thick, -latex] (env.west |- 0,-0.1) -- node[below]{$s_{t+1}$} (-2, -0.1);
        \draw[-latex] (-2, 0.1) -- ++(-1, 0) -- ++(0, 0.65) node[right]{Reward $r_t$} -- ++(0, 0.65) -- (a.west |- 0, 1.4);
        \draw[thick, -latex] (-2, -0.1) -- ++(-1.2, 0) -- ++(0, 0.85) node[left]{State $s_t$} -- ++(0, 0.85) -- (a.west |- 0, 1.6);
        \draw[thick, -latex] (a.east) -- ++(2.5, 0) -- ++(0, -0.75) node[right]{Action $a_t$} -- ++(0, -0.75) -- (env.east);
    \end{tikzpicture}
    \caption{Reinforcement learning feedback loop diagram.}
    \label{fig:feedback-loop}
\end{figure}
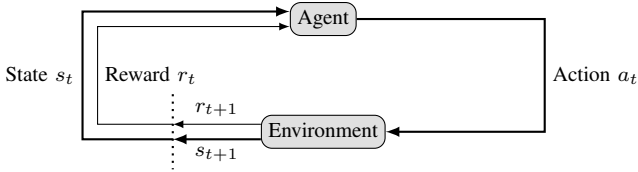

\subsection{Blockchain Technology}
Blockchain is a system designed for peer-to-peer networks that are decentralized, allowing for a secure and tamper-proof ledger maintained by the network participants themselves~\cite{15nakamoto2008bitcoin}. This contrasts with centralized systems; blockchain operates without a single point of control. It gained initial prominence with the launch of Bitcoin, the first cryptocurrency, and has since expanded its applications across various fields, such as finance, agriculture, health, and more. The structure of a blockchain can be thought of as a series of data blocks that are securely linked together using cryptographic principles. Each block contains a collection of transactions and is connected to the previous block via a cryptographic signature known as a hash. Should any alteration be attempted on a completed block, the hash will change, signalling a break in the integrity of the chain. Blocks are added to the blockchain through a consensus process, which often requires computational work to validate new entries. This process includes the use of a nonce, a number found by a network participant that when used in a hashing function, satisfies certain conditions set by the blockchain protocol. As the chain grows, altering any information retroactively becomes increasingly complex. Typically, each block includes certain information, such as a timestamp, its own unique identity, the hash of the previous block, a Merkle tree root which summarily represents the included transactions, and a nonce value, among other transaction details. This chained data structure ensures the fidelity and security of the transaction history, making blockchain a robust and trustworthy technology for recording transactions over time. In traditional blockchain systems, there is an inherent delay in processing transactions.

For blockchain technology, delays result from the time it takes for a transaction to be verified and added to a block, as well as by generating blocks and their corresponding arrival at all other nodes. The process involves multiple nodes in the network validating the blocks and transactions, which ensures security and decentralization but also introduces latency. In contrast, real-time blockchain aims to reduce these delays significantly, offering a solution where transactions are processed and confirmed in a much shorter timeframe. This is achieved through various means, such as different consensus mechanisms, increased block generation speed, or off-chain transaction channels. Real-time blockchain is particularly beneficial for applications requiring fast and reliable transaction processing, like financial services, gaming, and IoT operations, where traditional blockchain delays could hinder performance and usability.

\section{Research Challenges}\label{sec:Research Challenges}
\subsection{Outline the Current State of Research}
The journey of task offloading strategies in fog computing has been marked by a continuous search for optimizing key performance metrics, such as delay, energy consumption, security, and cost efficiency. Initial approaches focused on delay minimization through innovative algorithms like Exact Solutions and game-theoretic models, aiming to reduce latency for delay-sensitive tasks. Strategies evolved to address energy efficiency, with proposals ranging from incentive-based cloud-IoT offloading schemes to software-defined networks (SDN) based architectures, enhancing flexibility and decision-making in offloading policies. As the complexity of fog environments and the diversity of IoT applications grew, the focus expanded to delay, security, and energy considerations, adopting algorithms that could dynamically balance these critical factors. Techniques, such as partial task offloading and energy-aware scheduling emerged, incorporating more exacted decision-making frameworks to cater to the specific requirements of varied applications, from vehicular fog computing to health-care. The reliability and cost efficiency of offloading decisions also gained prominence, with strategies developing to ensure that fog computing architectures could support increasingly demanding applications without compromising on service quality or operational costs~\cite{16aazam2018offloading}. This broadened the scope of task offloading strategies to include considerations like resource allocation prediction, task scheduling, service latency, and quality loss trade-offs, pushing towards more adaptable solutions. Figure~2 represents the concept of task offloading in a Fog-Computing computing architecture involving IoT devices, fog computing nodes, and cloud servers. The process of task offloading is meticulously designed to streamline the interaction between IoT devices and the multi-layered architecture. This process begins at the IoT layer, where end devices, embedded with sensors and other local computing resources, initially generate tasks. These tasks, depending on their complexity and the immediate computational capacity available at the edge, may require offloading to more capable layers for efficient processing. In the next step, fog nodes assess the incoming tasks for their computational and storage needs and decide whether to process them within this layer or forward them further. This decision-making is critical and is based on factors, such as the task’s requirements, the available resources, and the desired efficiency in terms of time and energy consumption. For tasks that are either too demanding for the fog layer or optimized for centralized processing, the final offloading destination is the cloud layer. This layer, characterized by its vast computational and storage capabilities, is equipped to handle high-demand tasks offloaded from the lower layers. The virtual cloud server, supported by an underlying physical server infrastructure, ensures that these tasks are executed efficiently, leveraging the cloud’s resources. This offloading mechanism ensures that tasks generated by IoT devices are processed in the most appropriate layer, optimising resource utilization across the system and enhancing overall performance and security also~\cite{17tran2022survey}.By dynamically allocating tasks based on their needs and the available resources at each layer, the architecture supports a flexible and efficient processing model.

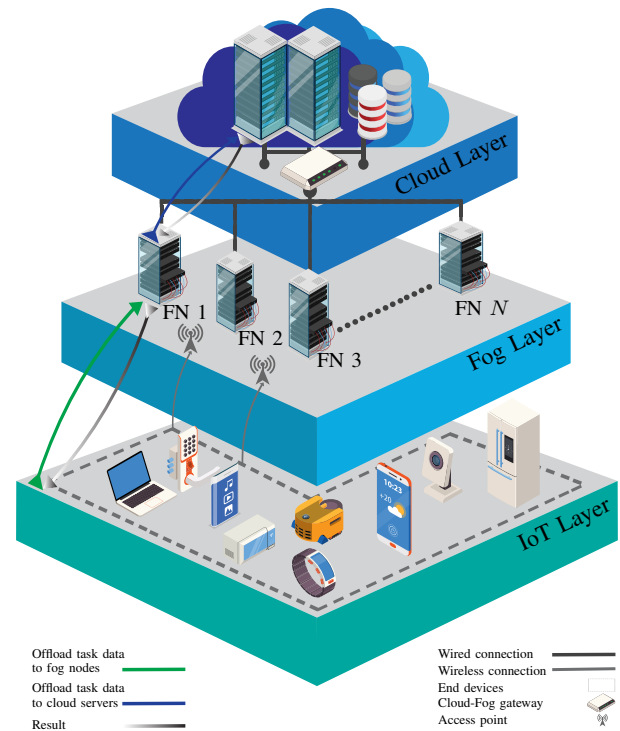
\begin{figure}[!h]
    \centering
    \small
    \begin{tikzpicture}
        \node at (0, 0) {\includegraphics[width=8cm]{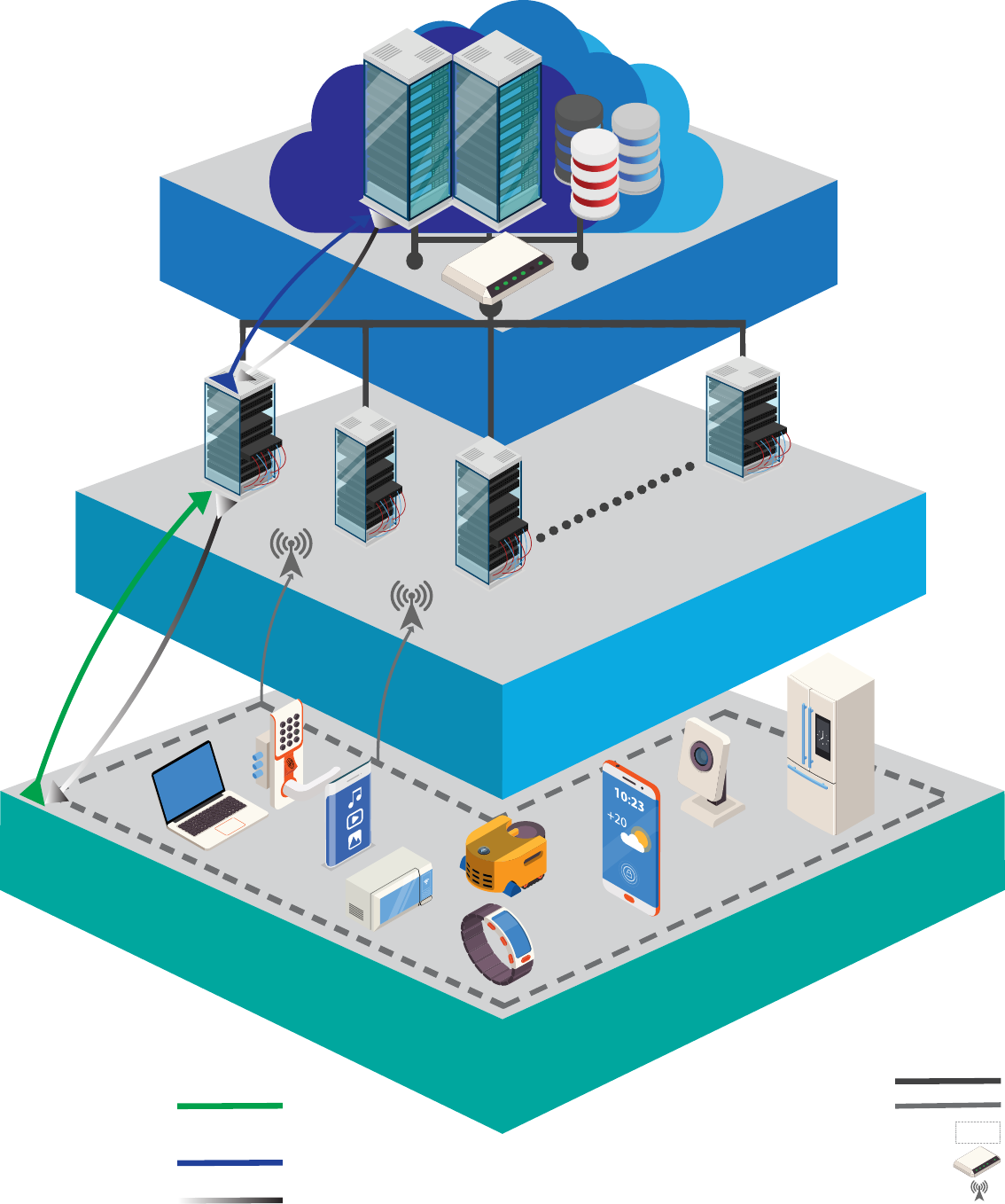}};
        \node[rotate=25] at (1.8, 2.7) {Cloud Layer};
        \node[rotate=25] at (2.65, 0.3) {Fog Layer};
        \node[rotate=25] at (3.3, -2.1) {IoT Layer};
        \node at (-1.75, 0.75) {\footnotesize FN 1};
        \node at (-0.75, 0.4) {\footnotesize FN 2};
        \node at (0.3, 0.1) {\footnotesize FN 3};
        \node at (2.2, 0.8) {\footnotesize FN $N$};
        \node[anchor=west] at (-3.9, -3.8) {\tiny Offload task data};
        \node[anchor=west] at (-3.9, -4) {\tiny to fog nodes};
        \node[anchor=west] at (-3.9, -4.25) {\tiny Offload task data};
        \node[anchor=west] at (-3.9, -4.45) {\tiny to cloud servers};
        \node[anchor=west] at (-3.9, -4.75) {\tiny Result};
        \node[anchor=west] at (1.5, -3.8) {\tiny Wired connection};
        \node[anchor=west] at (1.5, -4.025) {\tiny Wireless connection};
        \node[anchor=west] at (1.5, -4.25) {\tiny End devices};
        \node[anchor=west] at (1.5, -4.475) {\tiny Cloud-Fog gateway};
        \node[anchor=west] at (1.5, -4.7) {\tiny Access point};
    \end{tikzpicture}
    \caption{The architecture of task offloading in a fog computing environment.}
    \label{fig:architecture}
\end{figure}

Kishor et al.~\cite{18kishor2022task} have employed metaheuristic algorithms, such as the Smart Ant Colony Optimization (SACO), to facilitate efficient task offloading. These algorithms draw inspiration from natural processes and have shown considerable promise in optimising resource allocation and minimizing latency, thereby enhancing the Quality of Service (QoS) in IoT applications. The SACO algorithm, in particular, has demonstrated its effectiveness by significantly reducing task offloading time compared to traditional methods, such as Round Robin and throttled scheduler algorithms. By mimicking the foraging behaviour of ants, SACO efficiently distributes tasks across fog nodes, ensuring optimal utilization of resources and timely data processing. It becomes evident that such innovative offloading strategies are pivotal in overcoming the limitations of cloud computing in the context of real-time, sensor-based applications. Jiang et al.~\cite{19jiang2018delay} introduced a delay-aware task offloading scheme for shared fog networks, aiming to efficiently schedule tasks with varying delay sensitivities. A mathematical model is developed to represent fog networks, with a solution method based on problem-specific analysis. Simulations show the effectiveness of the proposed scheme, removing impractical assumptions from previous works and offering insights for improved task offloading in fog computing. The article also explores the balance between efficiency and fairness in optimization problems in cloud and edge computing, considering different objective functions like minimizing task inefficiency. The authors reference related works and focus on resource management for networked systems, cloud/edge computing, and big data systems in their research. Ke et al.~\cite{20ke2021deep} introduced a priority-aware task offloading scheme in vehicular fog computing using DRL. This scheme encourages vehicles to share their idle computing resources through dynamic pricing, taking into account task priority, service availability, and vehicle mobility. The problem is framed as a Markov decision process, with a soft actor-critic based DRL algorithm developed to maximize utility. Extensive simulations confirm the effectiveness of the proposed scheme over traditional algorithms. The study by Shi et al.~\cite{21shi2020priority} specifically focuses on priority-aware task offloading in vehicular fog computing, comparing the proposed algorithm with random, greedy-based algorithms, and Double Deep Q-Network. Results demonstrate that the proposed algorithm surpasses the others in mean utility, task completion ratio, and average delay. It ensures high-priority tasks are completed first and performs better in task completion and offloading delay. Overall, the study validates the efficiency of the proposed algorithm in dynamic vehicular environments.
In summary, the current state of task offloading in fog computing is characterized by a blend of energy-efficient algorithms, customizable offloading strategies, and innovative incentive mechanisms. These approaches collectively aim to enhance the capabilities of fog computing, addressing the diverse and evolving needs of IoT and mobile environments.

\begin{table*}[htb]
    \centering
    \caption{TABLE I.Comparison of task offloading methods in fog computing.}
    \begin{tabular}{C{3cm}*{3}{C{4.33cm}}}
        \toprule
        Criteria/Method & \textbf{DRL}~\cite{21shi2020priority} & \textbf{Metaheuristic Methods}~\cite{18kishor2022task} & \textbf{Exact Solutions}~\cite{19jiang2018delay}\\
        \toprule
        \textbf{Efficiency} & \multicolumn{3}{c}{}\\[2pt]
        \rowcolor{black!15}Computational Speed & High due to adaptive learning & High & Low\\
        Scalability & Excellent, adapts to large-scale environments & Good, but may require adjustments for scale & Very limited\\
        \rowcolor{black!15}Resource Utilization & Optimized through continuous learning & Better optimized than heuristics but varies & Highly optimized but impractical for large systems\\
        Energy Consumption & Reduced through efficient offloading decisions & Lower than heuristics but not as efficient as DRL & Optimized but at the cost of computational resources\\
        \midrule
        \textbf{Security} & \multicolumn{3}{c}{}\\[2pt]
        \rowcolor{black!15}Data Privacy & Enhanced by learning optimal offloading without exposing data & Better than heuristic but less than DRL & High, but often not the focus of design\\
        Attack Resistance & Improved through dynamic adaptation to threats & Better through diversity of solutions but slower to adapt & High for known threats, low for new threats\\
        \rowcolor{black!15}System Integrity & Maintained through continuous monitoring and adaptation & Good, with potential for periodic updates & High, but static and may be bypassed over time\\
        Authentication \& Access Control & Advanced, can integrate with state-of-the-art mechanisms & Moderate to high, depending on the method & High, but rigid and may not adapt well to new access patterns\\
        \toprule
        \textbf{Overall Performance} & Superior due to adaptability, learning capabilities, and ability to optimize for multiple objectives simultaneously & Very good, offering a balance between solution quality and computational effort, but can be unpredictable & Excellent in terms of achieving optimal solutions but at the cost of practicality in dynamic or large-scale environments\\
        \bottomrule
    \end{tabular}
    \label{tab:comparison-task-offloading-methods}
\end{table*}

\subsection{The Limitations of Existing Task Offloading Mechanisms}
In the evolving landscape of fog computing, task offloading presents a spectrum of security challenges. While fog computing ostensibly enhances security by minimizing reliance on centralized storage and extensive Internet connectivity, it inherits a suite of vulnerabilities from cloud computing. This content tries to resolve a critical examination of potential security risks inherent in task offloading processes. One of the typical security vulnerabilities arises from the physical and operational remoteness of cloud services, which fog computing seeks to ameliorate~\cite{22elgendy2020efficient}. Despite this, the transference of tasks to Fog Nodes (FNs) introduces complexities in safeguarding data integrity and confidentiality. The main point of the issue lies in the inherent limitations of FNs – their constrained computational resources and diminutive stature complicate the execution of robust security algorithms essential for mitigating threats like man-in-the-middle attacks, eavesdropping, and denial-of-service attacks. Moreover, the application of security measures, though imperative, adds to the energy demands of these devices, which leads to the necessity to make a compromise between required security and operational efficiency. This is further complicated by the latency issues arising from cryptographic operations, as edge devices, often comprised of small-scale servers, struggle with timely data encryption, for instance, thus increasing the latency within the network. The integration of middleware IoT security solutions propose a bridge between cloud and fog computing; however, vulnerabilities remain, especially in scenarios involving session resumption algorithms. These algorithms, designed for efficiency, could potentially be exploited by attackers to hijack sessions, suggesting a pressing need for improvement in secure session management. While new methods cover the security and privacy concerns within fog computing, there exists a palpable gap in addressing the concurrent optimization of delay, energy consumption, and security. The dynamic nature of fog networks, where nodes can seamlessly join or exit, further complicates the security paradigm, necessitating novel approaches to ensure the integrity and privacy of these fluid systems. Leveraging machine learning algorithms for attack detection in fog environments represents a promising frontier. These algorithms could potentially enhance data security and processing by identifying and mitigating threats in real-time. However, the practical implementation of such solutions is restricted by the limited computational resources of FNs, underscoring the need for innovative solutions that balance security, efficiency, and resource constraints. The task offloading process introduces a new set of challenges that require comprehensive strategies to address. Future research should aim at developing solutions that not only secure the fog computing environment but also optimize performance metrics, such as delay and energy consumption, thereby ensuring a secure, efficient, and resilient fog computing ecosystem.
In the realm of computational offloading, the imperative for robust security frameworks encompasses a dual-faceted approach. Firstly, it necessitates the establishment of mechanisms, such as confidentiality, integrity, availability, access control, and authentication. These measures are pivotal in safeguarding the communication between IoT devices and fog computing (FC) servers, thus ensuring the protected execution of computation offloading processes. The challenges posed in this domain often mirror those encountered within cloud computing; however, the unique characteristics of FC, including the limited resources of IoT devices and the reliance on wireless access, exacerbate the complexity of implementing effective security solutions. The second facet positions the FC server as a bulwark for the security of IoT devices, recognizing that these resource-constrained entities are often ill-equipped to support advanced security algorithms, such as group signatures. This realization prompts a paradigm where the security functionalities traditionally resident on IoT devices are instead offloaded to EC servers, which then assume the role of executing these tasks on behalf of the IoT devices. This shift, while pragmatic, introduces a spectrum of security considerations necessitated by the heterogeneity of IoT devices. This diversity encompasses varying communication standards, dynamic security configurations, and the constant evolution of security threats, thereby mandating a multifaceted and comprehensive approach to security within the FC ecosystem. Merging this perspective with the earlier discussion on fog computing and task offloading illuminates the broader spectrum of security risks and challenges across different computing paradigms. The intersection of fog computing and FC delineates a complex landscape where the task of securing offloaded computations becomes increasingly intricate. The offloading of computational tasks or security functions to FC servers, while pragmatic, open a torrent of privacy concerns, arguably more daunting than those faced in cloud computing environments. In addressing privacy concerns within fog computing, the application of Oblivious Random-Access Machine (ORAM)\cite{23YUAN20181} techniques, traditionally used to obscure user access patterns in cloud environments, is crucial. However, to maintain the latency advantages of fog computing, it can be mentioned the ThinORAM scheme ~\cite{24huang2019thinoram}, emerges as a tailored solution. ThinORAM adapts ORAM for fog computing, effectively balancing performance, security, and privacy without the significant drawbacks of increased energy consumption, latency, and computational overheads. This approach not only aligns with the operational dynamics of fog computing but also sets a precedent for future research to develop security solutions that navigate the trade-offs between security, privacy, and system performance, fostering a secure, efficient, and resilient distributed computing ecosystem.

\subsection{The Potential of DRL to Surmount These Limitations}
In fog computing, task offloading mechanisms face various limitations, particularly in terms of security. These mechanisms often struggle with ensuring data privacy, maintaining integrity, and preventing unauthorized access, as fog nodes are typically distributed and closer to end-users, increasing vulnerability to attacks. Moreover, resource management, latency, and network bandwidth are additional challenges that impact the efficiency and reliability of offloading tasks in fog environments. Deep Reinforcement Learning (DRL) presents a promising solution to overcome these challenges by enabling adaptive and intelligent decision-making in dynamic and uncertain environments. DRL can optimize resource allocation, improve task scheduling, and enhance security measures through its ability to learn and adapt from the behaviour of the system and threats. However, there is a significant research gap in fully exploiting the potential of DRL for security enhancement in fog computing. While DRL can potentially address issues like anomaly detection and response to evolving threats, more research is needed to develop robust DRL models that are specifically tailored for the unique challenges of fog computing environments, ensuring they are effective against a wide range of security threats while also optimising computational efficiency.

\subsection{The Potential of Combining DRL with Blockchains}
Combining DRL with blockchain technology could further secure task offloading by creating a decentralized and transparent ledger, reducing risks of tampering and ensuring data integrity, thus boosting overall efficiency and reliability in fog computing environments.
In fog computing, the necessity for immutable storage stems from the need to ensure data integrity and prevent unauthorized alterations. This is vital for maintaining trust in distributed computing environments, where data is frequently offloaded and processed across various nodes. Immutable storage guarantees that once data is recorded, it remains unchanged, providing a reliable foundation for decision-making processes and system operations, and safeguarding against data breaches or manipulations.

Further research challenges concern security considerations with blockchain integration.
While blockchain promises enhanced security and immutability, its integration into fog computing for task offloading raises critical security questions. The inherent complexity and new interfaces introduced by blockchain can potentially open up new vulnerabilities or exacerbate existing ones. It is imperative to scrutinize how security mechanisms of blockchains align with the unique demands and threat models of fog computing environments. This scrutiny is crucial to ensure that the solution does not inadvertently compromise the very security it aims to bolster.

The application of blockchain in fog computing is not without its efficiency challenges. The nature of blockchains, characterized by slower transaction speeds and block generation times, poses significant questions regarding its suitability for task offloading scenarios, which require rapid response times between the fog layer and IoT devices. The feasibility of achieving time-predictable transactions with blockchain is a concern, given the latency-sensitive nature of many fog computing applications. While blockchain technology presents a promising solution for enhancing security and integrity in fog computing environments, it is vital to optimize and adapt its application to meet the specific efficiency benchmarks required for effective task offloading. This approach underscores the potential of blockchain as a key technology in fog computing, provided that its implementation is fine-tuned to align with the unique demands of this context.

\section{Proposed Solution}
\subsection{Overview of the Proposed Solution}
Cloud computing's security risks stem from its centralized data handling and the physical gap from users' devices. Fog computing, a complementary approach  that situates computing resources closer to the data source or edge of the network, offers improved security by reducing reliance on Internet connectivity for data processing and storage. Despite these advantages, fog computing is not without its challenges. It inherits some of the security risks of cloud computing and introduces new ones due to the limited capabilities and resources of FNs, which can affect task offloading and security algorithm execution. Implementing security measures in fog computing can also lead to increased energy consumption and latency due to the encryption demands on smaller-scale servers. A middleware solution is proposed in this article that can maintain security as much as possible at the same time as it does not reduce the task offloading speed through blockchain. However, we are aware of this matter that future research in fog computing security could should focus on more optimising the balance between delay, energy consumption, and security. 
To address the challenge of enhancing system robustness and security against phishing attempts within fog computing environments, it is pivotal to recognize the complexities introduced by the vast data generated by interconnected IoT devices. These devices, operating within a smart ecosystem, contribute to a data-intensive environment that necessitates efficient offloading to fog or cloud layers for subsequent processing and storage. The critical concern arises when these devices, in an attempt to offload data, inadvertently direct their computations or sensitive data to compromised neighbouring fog servers. Such incidents, often resulting from sophisticated phishing attacks, underscore the vulnerability of the system and spotlight the imperative need for fortified security measures. This situation not only raises significant security concerns but also adversely affects network performance and energy efficiency, particularly when dealing with the intricacies of managing overloaded fog computing nodes. The interplay between ensuring robust security protocols and maintaining optimal network performance becomes increasingly complex, however , a of the  promising avenue for enhancing system security lies in the integration of blockchain technology~\cite{25jain2022blockchain}~\cite{26alam2023deep}.
The inherent blockchain characteristics of decentralization, transparency, and immutability present a novel approach to securing data transactions across the network. By leveraging blockchain technology within the fog computing paradigm, it is possible to establish a secure, tamper-proof system for data exchange and processing. This approach not only mitigates the risk of unauthorized access and phishing attacks but also contributes to the overall efficiency and reliability of the network infrastructure. This integration is anticipated to address the prevailing security challenges effectively, thereby enhancing the resilience and trustworthiness of the system against potential cyber threats.  Integrating an efficiency enhancement algorithm with blockchain within a fog computing environment involves a sophisticated coordination mechanism that leverages both technologies to optimize performance and security simultaneously. Imagine an algorithm that dynamically adjusts the distribution of tasks based on near real-time network conditions and device capabilities, while also ensuring data integrity and security through blockchain’s decentralized ledger. This algorithm operates continuously, analysing the state of the network, including workload distribution, device energy levels, and current latency metrics. The synergy between the dynamic efficiency enhancement algorithm and the blockchain ensures not only optimal resource utilization but also robust security, creating a resilient and adaptable fog computing ecosystem.
Blockchain ensures that data related to task offloading is securely stored and remains unaltered. This integrity is crucial for sensitive applications, ensuring that the offloaded tasks and their outcomes are reliable and trustworthy. In line with this, Figure 3 illustrates a diagram of task offloading strategy in fog computing environment. This strategy integrates blockchain technology to enhance system security~\cite{27xu2019edge}.
Blockchain's immutability and transparency are key features that stand out for task offloading processes. Once a transaction is recorded on a blockchain, it cannot be altered, providing a secure and trustworthy audit trail. This immutability ensures that the record of task offloading decisions and actions is preserved accurately, fostering trust among participants. Moreover, the transparency inherent in blockchain allows all network participants to view and verify transaction histories, ensuring the integrity and verifiability of the task offloading process. Furthermore, blockchain offers advantages over other immutable storage solutions through features like smart contracts, which automate and enforce task offloading processes without human intervention. It must be mentioned that the concerns raised regarding the efficiency and response times of blockchain technology in the context of task offloading between the fog layer and the IoT layer are valid. However, our research aligns with recent advancements in the field. Notably, a study by Lee et al. ~\cite{28lee2023rt} demonstrates a novel approach to integrating real-time scheduling principles into blockchain systems, aiming to ensure time-sensitive transactions. This method addresses the critical challenge of achieving time-predictable transactions in blockchain. By modifying the blockchain architecture to preferentially select transactions with the earliest deadlines they have shown that it is possible to meet the stringent response time requirements essential for efficient task offloading in fog computing environments. Their work provides a promising direction for enhancing the efficiency and predictability of blockchain transactions, which is pivotal for our research on task offloading in fog computing with deep reinforcement learning.

\begin{figure}[tb!]
    \centering
    \footnotesize
    \begin{tikzpicture}
        \node[draw, fill=black!12, anchor=west] (drl) at (0, 0) {DRL Agent};
        \node[draw, fill=black!12, anchor=east] (bc) at (8, 0) {Blockchain\vphantom{j}};
        \node[draw, fill=black!12] at (4.5, 0) (env) {Environment\vphantom{j}};
        \draw (drl.south) -- ++(0, -5);
        \draw (env.south) -- ++(0, -5);
        \draw (bc.south) -- ++(0, -5);
        \node[draw, fill=yellow!12] at (2.625, -0.6) {Action: Offloading decision (FN/cloud)};
        \draw[-latex] (drl.south |- 0,-1.2) -- node[above]{Perform action $a_t$} (env.south |- 0,-1.2);
        \draw[dotted, -latex] (env.south |- 0,-1.8) -- node[above]{Result of action} node[below]{(success/failure)} (drl.south |- 0,-1.8);
        \node[draw, fill=yellow!12] at (5.9, -2.2) {State: current system state};
        \draw[-latex] (env.south |- 0,-2.8) -- node[above]{Provide new state $s_t$} (drl.south |- 0,-2.8);
        \node[draw, fill=yellow!12, text width=6.8cm, text centered] at (4, -3.2) {Reward: based on action outcome};
        \draw[-latex] (env.south |- 0,-3.8) -- node[above]{Provide reward $r_t$} (drl.south |- 0,-3.8);
        \draw[-latex] (drl.south |- 0,-4.4) -- node[above]{Record action \& outcome on Blockchain} (bc.south |- 0,-4.4);
        \draw[dotted, -latex] (bc.south |- 0,-5) -- node[above]{Confirm transaction integrity} (drl.south |- 0,-5);
    \end{tikzpicture}
    \caption{Sequence diagram of the task offloading strategy using DRL in fog computing, incorporating blockchain technology.}
    \label{fig:sequence-diagram}
\end{figure}
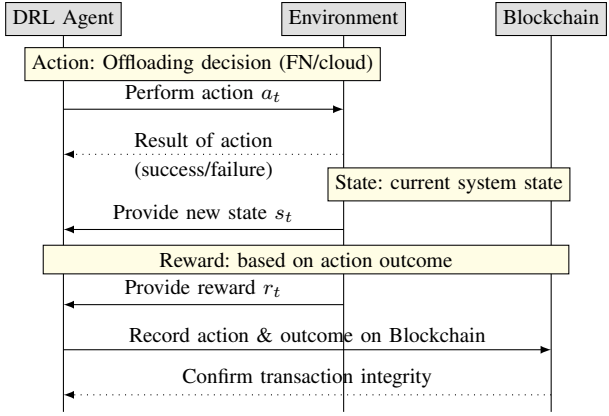

In the following section, we delineate the process of our proposed method step by step : 
We know in the first stage to optimize the efficiency of fog computing, the task offloading model must be rigorously defined. Tasks are characterized by parameters akin to those in blockchain transactions within the RT-Blockchain system~\cite{28lee2023rt}:inter-arrival time ($T_i$), relative deadline ($D_i$), and task size ($S_i$). Each task, akin to a blockchain transaction, undergoes a process of validation and scheduling for execution, maintaining the integrity of the fog computing framework.

A sophisticated task offloading algorithm in fog computing considers the translation of user-level task parameters to the more granular slot-level parameters for fog nodes. For instance, given user-level parameters ($T_i$, $D_i$, $S_i$), the slot-level parameters can be calculated taking into account network latencies and computational resources, ensuring the task can be scheduled effectively within the operational constraints.

\subsection{Deep Reinforcement Learning Integration}
Machine learning algorithms present a potential method for improving attack detection in fog environments, but their application is constrained by the limited resources of FNs.The use of DRL within this model stands to many improvements how tasks are offloaded in fog environments. By dynamically learning the optimal policy for task offloading based on state ($S$), action ($A$), and reward ($R$), the DRL agent responds to the complexities of the network. This optimization is grounded in a reward function, $R(S, A)$, which motivates actions that minimize latency and maximize resource utilization while ensuring security.

\subsection{Time-Predictable Transaction Framework Adaptation}
Adapting the time-predictable transaction framework involves setting the upper bounds for fog computing task processing ($C_{gen}^{fog}$) and validation time ($C_{val}^{fog}$), paralleling the blockchain model. This adaptation ensures that tasks are offloaded and processed within the constraints of fog computing, optimizing both security and efficiency without compromising the predictability of the system:
\begin{equation}
C_{gen}^{fog} = \alpha \times C_{gen}^{block}
\end{equation}
\begin{equation}
C_{val}^{fog} = \beta \times C_{val}^{block}
\end{equation}
Where $\alpha$ and $\beta$ are scaling factors that adjust the blockchain constants $C_{gen}^{block}$ and $C_{val}^{block}$ for fog computing realities.

\subsection{Demand Bound Function and Load in Fog Computing}
Incorporating the Demand Bound Function (DBF) from the blockchain model into fog computing ensures that the system load does not exceed its capacity. The adapted DBF for fog computing is defined as:
\begin{equation}
DBF_i(\Delta) = \max(0, \left\lceil \frac{\Delta - (D_i - T_i)}{T_i} \right\rceil \times C_i)
\end{equation}
This equation calculates the cumulative demand that tasks impose on the fog computing resources within a specific interval ($\Delta$), ensuring that the system load remains within capacity and tasks are completed within their deadlines:
\begin{equation}
\text{Load}(\Delta) = \frac{1}{\Delta} \sum_{i=1}^{n} DBF_i(\Delta)
\end{equation}

\subsection{Schedulability and Security Analysis}
We propose a method to evaluate the schedulability of tasks in fog computing that parallels blockchain's validation theorems. This analysis ensures that offloaded tasks are feasible within the deadlines and system capacities, thereby enhancing security.

The DRL model will include security considerations, learning to recognize and mitigate potential threats. This model will be formulated to optimize not just for efficiency but also for robust security measures, such as validating task authenticity and preventing overloading of nodes.

\subsection{Algorithm and Evaluation}
The algorithm that combines DRL with the time-predictable task offloading framework will be evaluated on metrics, such as efficiency, security, and adherence to time constraints. Evaluation will incorporate real-time data and the following predictive formula for schedulability:
\begin{equation}
\text{Schedulability} = \frac{\sum \text{Completed Tasks}}{\sum \text{Scheduled Tasks}}
\end{equation}
While Formula (5) provides a straightforward metric for evaluating the efficiency of our task offloading strategy by comparing the number of completed tasks to the total scheduled tasks, it's crucial to understand the underlying factors contributing to uncompleted tasks. These may include network latency, which delays task execution, resource constraints that prevent tasks from being processed, security protocols that interrupt task execution for safety reasons, or other operational inefficiencies. By analyzing these factors in depth, we can gain more insights into the system's performance and identify targeted improvements for our fog computing solution

\section{Conclusion and Future Work}\label{sec:Conclusion and Future Work}
In conclusion, this study's exploration into the integration of Deep Reinforcement Learning (DRL) and blockchain technology within fog computing environments not only reveals critical insights but also opens new avenues for future research. While our findings underscore the potential of this integration to enhance security and efficiency in task offloading, they also highlight the need for further optimization of blockchain technology to meet the specific demands of fog computing. We observed that DRL's effectiveness in dynamic decision-making is significantly influenced by the availability and quality of training data. Moreover, current blockchain implementations face challenges like transaction speed and resource consumption that could affect their suitability for time-sensitive fog computing applications.
Future research should delve into these challenges, seeking more scalable and efficient blockchain solutions and refining DRL models for better adaptation to fog computing's complexity. Specifically, exploring heuristic approaches like Fuzzy Reinforcement Learning could provide valuable insights into handling uncertainty in decision-making processes, an inherent aspect of fog computing environments. Additionally, investigating other heuristics, such as genetic algorithms and swarm intelligence, could offer alternative strategies for optimizing task offloading and resource allocation, further enhancing the adaptability and performance of fog computing systems.

\renewcommand*{\bibfont}{\footnotesize} 
\setlength{\labelnumberwidth}{0.45cm}
\printbibliography

\end{document}